# Probability-free foundation of continuum mechanics equations irreversibility: connection with particle dynamics

Victor V. Zubkov[1,2]

[1]Department of General Physics, Tver State University, Tver, 170002, Russia

[2] Department of General and Experimental Physics, Yaroslav-the-Wise Novgorod State University, Veliky Novgorod, 173003, Russia

An equation describing the irreversible evolution of the local density of a continuous medium without involving any statistical hypotheses and assumptions is derived. The derivation is based on the smoothing of the microscopic dynamic characteristics of a many-body system, taking into account the retardation of the interactions between them. The resulting equation generalizes the classical equation of motion for fluids. Several particular cases of the resulting equation, as well as its connection with the dynamic density functional theory are considered.

## 1. Introduction

Mechanics of continuous media (MCM), as a rule, is justified within the framework of statistical physics [1-3]. Statistical physics, in turn, is based on classical mechanics, but despite this fact, there is generally no consistent connection between Newtonian mechanics and MCM. There are several reasons for this. First, the statistical physics necessarily requires the Hamiltonian formulation of classical mechanics. Within the framework of Hamiltonian mechanics, it is impossible to describe any motion of the system, because not any mechanical system can be described by Hamiltonian. Secondly, the methods of thermodynamics are used when creating models within the framework of MCM. Thermodynamics is the axiomatic science, describing equilibrium systems or weakly nonequilibrium ones. At the same time, the nonequilibrium processes described with the help of entropy concept cannot be explained from the point of view of classical mechanics. Thirdly, some concepts of MCM are not quite clear from physical point of view. For example, how to determine the density at a point? On the one hand, density is the derivative of mass with respect to volume; on the other hand, it is the integral of the particle distribution function over the momentum space of the system. Moreover, the density at a point, as a purely mathematical abstraction, generally contradicts the physical concept of density, which requires a representation of a physically small volume.

---

[1] Electronic mail: victor.v.zubkov@gmail.com.

There are attempts to explain irreversibility within the concept of deterministic chaos [4]. However, they do not represent a completely satisfactory solution to the problem of irreversibility, as well as attempts to explain irreversibility with the help of the hypothesis of nonuniqueness of the solution of the Cauchy problem for dynamical equations. Also a widely widespread point of view is the one set forth in the Penrose's book. [5]. Penrose [5] assumed that the second law of thermodynamics cannot be a simple consequence of the laws of classical mechanics. Rather, on the contrary, following Penrose, the second law must be attached to the laws of mechanics. In his opinion, this is the only way to describe irreversible processes. And further Penrose argues that "no amount of further stirring will convert the purple colourback to the original separated regions of red and blue, despite the time-reversibility of the submicroscopic physical processes underlying the mixing." [5, p.16]. However, recently it was shown in Ref. [6] that the equation of motion for the microscopic density of a system of particles becomes irreversible when taking into account the *actually existing retardation effect* of interaction between particles [7,8]. This means the possibility of a microscopic foundation of thermodynamics, as well as the possibility of obtaining the irreversible equations of motion of a continuous medium directly from Newtonian mechanics without involving any probabilistic assumptions and hypotheses. For this it is possible, by solving Zakharov's equation [6], to smooth out the solution and to make the passage from microscopic functions to mesoscopic or macroscopic variables (depending on the scale of smoothing). However, it is not entirely clear how to do it in a concrete way. As noted by Zakharov, the smoothing operators of Steklov or Sobolev [9,10] can be useful along this path. But, firstly, the choice of the kernels of these operators from the physical point of view is not unambiguous. Secondly, it is not clear how to smooth out those terms of the equation of motion that contain a product of smoothed functions. Another method, described in this paper, is based on the original use of smoothed functions or, more precisely, on the use of the smoothing operator for additive functions of a microscopic system. This smoothing operator associates the macrostate of the continuous media with the microstate of the many-body system. This method, proposed by Kuz'menkov [11, 12], naturally allows us to construct a MCM based on Newtonian mechanics without involving any probability assumptions. In this case, the concepts underlying the MSS become more transparent from a physical point of view.

This paper is aimed on derivation and analysis of the equation of motion for the macroscopic density (smoothed microscopic density) of a classical system of interacting particles taking into account the retarded interactions.

## 2. The method of smoothing the microscopic dynamic functions and the equation of motion for the local density of the medium

According to [11, 12], each microscopic dynamic additive quantity $\chi_i$ corresponding to the microscopic density

$$\chi(\mathbf{r},t) = \sum_{i=1}^{N} \chi_i \delta(\mathbf{r}-\mathbf{r}_i(t)), \qquad (1)$$

can be associated with the corresponding field function $\psi(\mathbf{r},t)$ using the smoothing operator $\hat{S}$:

$$\psi(\mathbf{r},t) = \hat{S}[\chi(\mathbf{r},t)] \equiv \frac{1}{\Delta}\sum_{i=1}^{N(\mathbf{r},t)} \chi_i = \frac{1}{\Delta}\int_{\Delta} \chi(\mathbf{r}+\xi,t)d^3\xi. \qquad (2)$$

Here $\Delta$ is a physically infinitesimal volume whose center is at a point $\mathbf{r}$, $N(\mathbf{r},t)$ is that a number of particles in this volume. It is easy to see that the smoothing operation (2) reduces to the selection of those particles $N(\mathbf{r},t)$ from the whole set $N$ by those which fall into the selected volume $\Delta$. This means that the action of the operator $\hat{S}$ can be reduced to using of the so-called discontinuous integral of Dirichlet [13].

In fact, formula (2) is the definition of a "point" of a continuous medium in its physical sense. In order that the function $\psi(\mathbf{r},t)$ can be considered as sufficiently smooth one, it is necessary that its increment $|\psi(\mathbf{r}+\delta\mathbf{r},t)-\psi(\mathbf{r},t)|$ is small compared to $\psi(\mathbf{r},t)$. If the number of particles of radius $r_0$ at a "point" of a continuous medium is defined as $N(\mathbf{r},t) \approx (nr_0^3)^{-5/4}$ ($n$ is the average particle concentration), then using the functions (2) a unified description of both kinematic and gasdynamic processes is possible [14].

Using the relation (2) introduced above we can write the scalar field of mass density

$$\rho(\mathbf{r},t) = \frac{1}{\Delta}\sum_{i=1}^{N(\mathbf{r},t)} m_i = \frac{1}{\Delta}\int_{\Delta}\sum_{i=1}^{N} m_i\delta(\mathbf{r}+\xi-\mathbf{r}_i(t))d^3\xi \qquad (3)$$

and the scalar field of concentration (density of the number of particles, local density of the medium)

$$n(\mathbf{r},t) = \frac{1}{\Delta}\int_{\Delta}\sum_{i=1}^{N}\delta(\mathbf{r}+\xi-\mathbf{r}_i(t))d^3\xi . \qquad (4)$$

Using the Fourier representation for $\delta$-functions, we rewrite (4) in the form

$$n(\mathbf{r},t) = \frac{1}{(2\pi)^3}\int d^3\mathbf{k}\, e^{i\mathbf{k}\mathbf{r}} F(\mathbf{k})\tilde{n}(\mathbf{k},t), \qquad (5)$$

where $\tilde{n}(\mathbf{k},t)$ are the collective coordinates

$$\tilde{n}(\mathbf{k},t) = \sum_{i=1}^{N} e^{-i\mathbf{k}\mathbf{r}_i} \quad , \tag{6}$$

introduced by Bohm and Pines [15], and $F(\mathbf{k}) = \frac{1}{\Delta}\int_\Delta e^{-i\mathbf{k}\boldsymbol{\xi}} d^3\xi$. The case $F(\mathbf{k})=1$ corresponds to a system of $N$ particles without the division into physically infinitesimal volume [6].

The vector field of the mass flux density associated with the motion of particles of the medium with velocities $\mathbf{v}_i(t)$ is defined as

$$\mathbf{j}(\mathbf{r},t) = \frac{1}{\Delta}\int_\Delta \sum_{i=1}^{N} m_i \mathbf{v}_i(t) \delta(\mathbf{r}+\boldsymbol{\xi}-\mathbf{r}_i(t)) d^3\xi = \frac{1}{\Delta}\sum_{i=1}^{N(\mathbf{r},t)} m_i \mathbf{v}_i(t) = \mathbf{v}(\mathbf{r},t)\rho(\mathbf{r},t). \tag{7}$$

Here we introduced a vector field $\mathbf{v}(\mathbf{r},t)$, i.e. the velocity of the center of mass of a physically infinitesimal volume:

$$\mathbf{v}(\mathbf{r},t) = \sum_{i=1}^{N(\mathbf{r},t)} m_i \mathbf{v}_i(t) \Big/ \sum_{i=1}^{N(\mathbf{r},t)} m_i . \tag{8}$$

Further we shall consider the motion of particles of the same mass. In this case

$$\mathbf{j}(\mathbf{r},t) = m\mathbf{v}(\mathbf{r},t) n(\mathbf{r},t). \tag{9}$$

From (7), (4) and (9) the equation of continuity follows:

$$\frac{\partial n(\mathbf{r},t)}{\partial t} = \frac{\partial}{\partial t}\frac{1}{\Delta}\int_\Delta \sum_{i=1}^{N} \delta(\mathbf{r}+\boldsymbol{\xi}-\mathbf{r}_i(t)) d^3\xi =$$
$$= -\frac{\partial}{\partial \mathbf{r}}\frac{1}{\Delta}\int_\Delta \sum_{i=1}^{N} \mathbf{v}_i \delta(\mathbf{r}+\boldsymbol{\xi}-\mathbf{r}_i(t)) d^3\xi = -\frac{1}{m}\mathrm{div}\,\mathbf{j}(\mathbf{r},t). \tag{10}$$

Differentiating the flux density with respect to time, we obtain:

$$\frac{\partial j_\alpha(\mathbf{r},t)}{\partial t} = \frac{1}{\Delta}\frac{\partial}{\partial t}\int_\Delta \sum_{i=1}^{N} m \mathrm{v}_{i\alpha}(t) \delta(\mathbf{r}+\boldsymbol{\xi}-\mathbf{r}_i(t)) d^3\xi = \Phi_\alpha(\mathbf{r},t) - \frac{\partial \Pi_{\alpha\beta}(\mathbf{r},t)}{\partial x_\beta} . \tag{11}$$

Here $\Pi_{\alpha\beta}(\mathbf{r},t)$ is a tensor field of flux density

$$\Pi_{\alpha\beta}(\mathbf{r},t) = \frac{1}{\Delta}\int_\Delta \sum_{i=1}^{N} m \mathrm{v}_{i\alpha}(t) \mathrm{v}_{i\beta}(t) \delta(\mathbf{r}+\boldsymbol{\xi}-\mathbf{r}_i(t)) d^3\xi , \tag{12}$$

and $\Phi_\alpha(\mathbf{r},t)$ is volume density of the force

$$\Phi_\alpha(\mathbf{r},t) = \frac{1}{\Delta}\int_\Delta \sum_{i=1}^{N} F_{i\alpha}(\mathbf{r}_i(t), \mathbf{v}_i(t), t) \delta(\mathbf{r}+\boldsymbol{\xi}-\mathbf{r}_i(t)) d^3\xi . \tag{13}$$

From (10) and (11) the equation for the density of number of particles follows:

$$\frac{\partial^2 n(\mathbf{r},t)}{\partial t^2} = -\frac{1}{m}\frac{\partial \Phi_\alpha(\mathbf{r},t)}{\partial x_\alpha} + \frac{1}{m}\frac{\partial^2 \Pi_{\alpha\beta}(\mathbf{r},t)}{\partial x_\alpha \partial x_\beta} . \tag{14}$$

If we know the forces through which the particles interact with each other and with the external environment, then solving equation (14), we can find the evolution of the particle number density.

Equation (14) is obtained without involving any probabilistic assumptions. Moreover, all field functions in this equation are obtained by smoothing (after application of the smoothing operator $\hat{S}$) the microscopic density (1), which expresses a clear relationship between Newtonian point mechanics and MCM. It is also important to note that the expression of this connection does not require the involvement of methods of statistical physics.

Formally, equation (14) can also be obtained on the basis of the first equation of the BBGKY hierarchy. The using of the methods of statistical physics makes it possible, in some approximation, to express the first term on the right-hand side of Eq. (14) in terms of the Helmholtz free energy functional. Then equation (14) is the basis of the dynamic density functional theory (DDFT) used to model the evolution of both simple fluids and colloidal systems [16-18]. However, in this case, field functions are determined by integrating the distribution functions [16-18]. Therefore, field functions already have a different (probabilistic) meaning, and the logical (consistent) connection between Newtonian mechanics and MCM disappears. We emphasize, that the above derivation of equation (14) is based only on Newton's second law and does not contain any additional assumptions (including probabilistic hypotheses).

Further analysis of Eq. (14) is related to the calculation of the volume density of the force (13), taking into account the retardation of the interactions between the particles.

### 3. Equation of motion taking into account the retarded interactions

To describe the evolution of the local density of a continuous medium, we use the second Newton's law:

$$\dot{\mathbf{p}}_i = -\frac{\partial}{\partial \mathbf{r}_i} U(\mathbf{r}_i, t). \tag{15}$$

The potential at the point $\mathbf{r}_i$, taking into account the retarded interactions, has the form:

$$\begin{aligned} U(\mathbf{r}_i, t) &= \sum_{k \neq i}^{N} U(\mathbf{r}_i(t) - \mathbf{r}_k(t - \tau_{ik})) \\ &= \sum_{k \neq i}^{N} \int U(\mathbf{r}_i(t) - \mathbf{r}_k(t')) \delta\left(t - t' - \frac{|\mathbf{r}_i - \mathbf{r}_k|}{c}\right) dt', \end{aligned} \tag{16}$$

where $\tau_{ik} = \dfrac{|\mathbf{r}_i - \mathbf{r}_k|}{c} \equiv \tau$ is the interactions retardation between $i$-th and $k$-th particles, $c$ is the light speed. It should be specially noted that the function $U(\mathbf{r})$ is defined for resting particles, while the function (16) depends not only on $\mathbf{r}$ and $t$, but also on the interactions retardation $\tau_{ik}$ between all the particles.

Taking into account (15) and (16), the expression for the volume density of force (13) can be represented in the form:

$$\begin{aligned}\mathbf{\Phi}(\mathbf{r},t) &= -\frac{1}{\Delta}\int_\Delta \sum_{i=1}^N \sum_{k\neq i}^N \delta(\mathbf{r}+\boldsymbol{\xi}-\mathbf{r}_i(t))\frac{\partial}{\partial \mathbf{r}_i}\int dt'\, U(\mathbf{r}_i(t)-\mathbf{r}_k(t'))\delta(t-t'-\tau)d^3\xi = \\ &= -\frac{1}{\Delta^2}\int dt' \int\limits_{\Delta(\mathbf{r})} d^3\xi \int\limits_{\Delta(\mathbf{r}')} d^3\xi' \int d^3\mathbf{r}'\, G(\mathbf{r}+\boldsymbol{\xi}-\mathbf{r}'-\boldsymbol{\xi}',t,t') \times \\ &\quad \times \sum_{i=1}^N \sum_{k\neq i}^N \delta(\mathbf{r}+\boldsymbol{\xi}-\mathbf{r}_i(t))\delta(\mathbf{r}'+\boldsymbol{\xi}'-\mathbf{r}_k(t')).\end{aligned} \quad (17)$$

Here

$$G(\mathbf{r}+\boldsymbol{\xi}-\mathbf{r}'-\boldsymbol{\xi}',t,t') = \frac{\partial}{\partial \mathbf{r}}\left[U(\mathbf{r}+\boldsymbol{\xi}-\mathbf{r}'-\boldsymbol{\xi}')\delta\left(t-t'-\frac{|\mathbf{r}+\boldsymbol{\xi}-\mathbf{r}'-\boldsymbol{\xi}'|}{c}\right)\right]. \quad (18)$$

After calculating the derivative in expression (18) and integrating in (17) with respect to time $t'$, we rewrite (17) in a slightly different form:

$$\begin{aligned}\mathbf{\Phi}(\mathbf{r},t) &= -\frac{1}{\Delta^2}\int\limits_{\Delta(\mathbf{r})} d^3\xi \int\limits_{\Delta(\mathbf{r}')} d^3\xi' \int d^3\mathbf{r}'\, G_1(\mathbf{r}+\boldsymbol{\xi}-\mathbf{r}'-\boldsymbol{\xi}') \times \\ &\quad \times \sum_{i=1}^N \sum_{k\neq i}^N \delta(\mathbf{r}+\boldsymbol{\xi}-\mathbf{r}_i(t))\delta(\mathbf{r}'+\boldsymbol{\xi}'-\mathbf{r}_k(t-\tilde{\tau})) - \\ &\quad - \frac{1}{\Delta^2}\int\limits_{\Delta(\mathbf{r})} d^3\xi \int\limits_{\Delta(\mathbf{r}')} d^3\xi' \int d^3\mathbf{r}'\, G_2(\mathbf{r}+\boldsymbol{\xi}-\mathbf{r}'-\boldsymbol{\xi}')\frac{\mathbf{p}'}{m} \times \\ &\quad \times \frac{\partial}{\partial \mathbf{r}'}\sum_{i=1}^N \sum_{k\neq i}^N \delta(\mathbf{r}'+\boldsymbol{\xi}'-\mathbf{r}_k(t-\tilde{\tau}))\delta(\mathbf{r}+\boldsymbol{\xi}-\mathbf{r}_i(t))\delta(\mathbf{p}'-\mathbf{p}_k(t-\tilde{\tau})).\end{aligned} \quad (19)$$

Here $G_1 = \dfrac{\partial U(\mathbf{r}+\boldsymbol{\xi}-\mathbf{r}'-\boldsymbol{\xi}')}{\partial \mathbf{r}}$, $G_2 = U(\mathbf{r}+\boldsymbol{\xi}-\mathbf{r}'-\boldsymbol{\xi}')\dfrac{1}{c}\dfrac{\mathbf{r}+\boldsymbol{\xi}-\mathbf{r}'-\boldsymbol{\xi}'}{|\mathbf{r}+\boldsymbol{\xi}-\mathbf{r}'-\boldsymbol{\xi}'|}$, and $\tilde{\tau}$ is the interactions retardation between points with coordinates $\mathbf{r}+\boldsymbol{\xi}$ и $\mathbf{r}'+\boldsymbol{\xi}'$:

$$\tilde{\tau} = \frac{|\mathbf{r}+\boldsymbol{\xi}-\mathbf{r}'-\boldsymbol{\xi}'|}{c}. \quad (20)$$

We expand the functions $G_1$ and $G_2$ in a series in powers of $\xi^\alpha - \xi'^\alpha$, and introduce the moments $S_1^{(\alpha_1\alpha_2\ldots\alpha_n)}(\mathbf{r},\mathbf{r}',t,t-\tilde{\tau})$ and $S_2^{(\alpha_1\alpha_2\ldots\alpha_n)}(\mathbf{r},\mathbf{r}',\mathbf{p}',t,t-\tilde{\tau})$:

$$S_1^{(\alpha_1\alpha_2...\alpha_n)}(\mathbf{r},\mathbf{r}',t,t-\tilde{\tau}) =$$
$$= \frac{1}{\Delta^2}\sum_{i=1}^{N}\sum_{k\neq i}^{N}\int_{\Delta(\mathbf{r})}d^3\xi\int_{\Delta(\mathbf{r}')}d^3\xi'\left(\xi^{\alpha_1}-\xi'^{\alpha_1}\right)\left(\xi^{\alpha_2}-\xi'^{\alpha_2}\right)...\left(\xi^{\alpha_n}-\xi'^{\alpha_n}\right)\times \quad (21)$$
$$\times\delta\left(\mathbf{r}+\boldsymbol{\xi}-\mathbf{r}_i(t)\right)\delta\left(\mathbf{r}'+\boldsymbol{\xi}'-\mathbf{r}_k(t-\tilde{\tau})\right).$$

$$S_2^{(\alpha_1\alpha_2...\alpha_n)}(\mathbf{r},\mathbf{r}',\mathbf{p}',t,t-\tilde{\tau}) =$$
$$= \frac{1}{\Delta^2}\sum_{i=1}^{N}\sum_{k\neq i}^{N}\int_{\Delta(\mathbf{r})}d^3\xi\int_{\Delta(\mathbf{r}')}d^3\xi'\left(\xi^{\alpha_1}-\xi'^{\alpha_1}\right)\left(\xi^{\alpha_2}-\xi'^{\alpha_2}\right)...\left(\xi^{\alpha_n}-\xi'^{\alpha_n}\right)\times \quad (22)$$
$$\delta\left(\mathbf{r}'+\boldsymbol{\xi}'-\mathbf{r}_k(t-\tilde{\tau})\right)\delta\left(\mathbf{r}+\boldsymbol{\xi}-\mathbf{r}_i(t)\right)\delta\left(\mathbf{p}'-\mathbf{p}_k(t-\tilde{\tau})\right).$$

Then the expression for the volume density of the force can be expressed in the final form:

$$\boldsymbol{\Phi}(\mathbf{r},t) = -\int d^3\mathbf{r}'\sum_{n=0}^{\infty}\frac{1}{n!}F^{(1)}_{(\alpha_1\alpha_2...\alpha_n)}S_1^{(\alpha_1\alpha_2...\alpha_n)}(\mathbf{r},\mathbf{r}',t,t-\tilde{\tau}) -$$
$$-\int d^3\mathbf{r}'\int d^3\mathbf{p}'\sum_{n=0}^{\infty}\frac{1}{n!}F^{(2)}_{(\alpha_1\alpha_2...\alpha_n)}\frac{\mathbf{p}'}{m}\frac{\partial}{\partial\mathbf{r}'}S_2^{(\alpha_1\alpha_2...\alpha_n)}(\mathbf{r},\mathbf{r}',\mathbf{p}',t,t-\tilde{\tau}). \quad (23)$$

Here

$$F^{(k)}_{(\alpha_1\alpha_2...\alpha_n)}(\mathbf{r}-\mathbf{r}') = \left.\frac{\partial^n G_k}{\partial x^{\alpha_1}\partial x^{\alpha_2}...\partial x^{\alpha_n}}\right|_{\xi^\alpha-\xi'^\alpha=0}. \quad (24)$$

Indices $\alpha$ in the expression (23) mean the summation. Thus, the equation for the local density of a continuous medium with taking into account the retarded interactions takes the final form:

$$\frac{\partial^2 n(\mathbf{r},t)}{\partial t^2} - \frac{1}{m}\nabla_\mathbf{r}\cdot\nabla_\mathbf{r}\boldsymbol{\Pi}(\mathbf{r},t) = \frac{1}{m}\nabla_\mathbf{r}\left(\int d^3\mathbf{r}'\sum_{n=0}^{\infty}\frac{1}{n!}F^{(1)}_{(\alpha_1\alpha_2...\alpha_n)}S_1^{(\alpha_1\alpha_2...\alpha_n)}(\mathbf{r},\mathbf{r}',t,t-\tilde{\tau}) + \right.$$
$$\left. +\int d^3\mathbf{r}'\int d^3\mathbf{p}'\sum_{n=0}^{\infty}\frac{1}{n!}F^{(2)}_{(\alpha_1\alpha_2...\alpha_n)}\frac{\mathbf{p}'}{m}\frac{\partial}{\partial\mathbf{r}'}S_2^{(\alpha_1\alpha_2...\alpha_n)}(\mathbf{r},\mathbf{r}',\mathbf{p}',t,t-\tilde{\tau})\right). \quad (25)$$

Performing the expansion of the moments in powers of the interactions retardation, it is not difficult to verify that after replacing

$$\mathbf{r} \to \mathbf{r},\ t \to -t,\ \mathbf{p} \to -\mathbf{p},$$

corresponding to inverse time operation, the left-hand side of (25) does not change, and the change in sign on the right-hand side occurs only for terms of the expansion with odd numbers. Consequently, the resulting equation is noninvariant with respect to time reversal and, as a consequence, describes the irreversible behavior of the system of the particles.

Let us suppose that the interaction potential is a smooth function, and it decreases quite rapidly with the distance between the particles. Then we can, after carrying out the expansion with respect to the interactions retardation, limit ourselves by the terms of the first order in $\tau$. We note that the following approximate relation is valid for small $\tau$:

$$\delta\left(\mathbf{r}'+\boldsymbol{\xi}'-\mathbf{r}_k(t-\tau)\right) \simeq \delta\left(\mathbf{r}'+\boldsymbol{\xi}'-\mathbf{r}_k(t)\right) + \tau\frac{\mathbf{p}_k(t)}{m}\frac{\partial}{\partial\mathbf{r}'}\delta\left(\mathbf{r}'+\boldsymbol{\xi}'-\mathbf{r}_k(t)\right). \quad (26)$$

Taking into account also that with a high degree of accuracy the interactions retardation $\tilde{\tau} = \frac{|\mathbf{r}+\boldsymbol{\xi}-\mathbf{r}'-\boldsymbol{\xi}'|}{c} \simeq \frac{|\mathbf{r}-\mathbf{r}'|}{c}$, we obtain

$$\Phi(\mathbf{r},t) = -\int d^3\mathbf{r}' \sum_{n=0}^{\infty} \frac{1}{n!} F^{(1)}_{(\alpha_1\alpha_2...\alpha_n)}(\mathbf{r}-\mathbf{r}') S_1^{(\alpha_1\alpha_2...\alpha_n)}(\mathbf{r},\mathbf{r}',t) - $$
$$-\int d^3\mathbf{r}' \int d^3\mathbf{p}' \sum_{n=0}^{\infty} \frac{1}{n!} \tilde{F}^{(2)}_{(\alpha_1\alpha_2...\alpha_n)}(\mathbf{r}-\mathbf{r}') \frac{\mathbf{p}'}{m} \frac{\partial}{\partial \mathbf{r}'} S_2^{(\alpha_1\alpha_2...\alpha_n)}(\mathbf{r},\mathbf{r}',\mathbf{p}',t). \qquad (27)$$

Here

$$\tilde{F}^{(2)}_{(\alpha_1\alpha_2...\alpha_n)}(\mathbf{r}-\mathbf{r}') = \frac{|\mathbf{r}-\mathbf{r}'|}{c} F^{(1)}_{(\alpha_1\alpha_2...\alpha_n)}(\mathbf{r}-\mathbf{r}') + F^{(2)}_{(\alpha_1\alpha_2...\alpha_n)}(\mathbf{r}-\mathbf{r}'). \qquad (28)$$

Then the equation for the local density in the mentioned approximation has the form:

$$\frac{\partial^2 n(\mathbf{r},t)}{\partial t^2} - \frac{1}{m} \nabla_\mathbf{r} \cdot \nabla_\mathbf{r} \Pi(\mathbf{r},t) = \frac{1}{m} \nabla_\mathbf{r} \left( \int d^3\mathbf{r}' \sum_{n=0}^{\infty} \frac{1}{n!} F^{(1)}_{(\alpha_1\alpha_2...\alpha_n)}(\mathbf{r}-\mathbf{r}') S_1^{(\alpha_1\alpha_2...\alpha_n)}(\mathbf{r},\mathbf{r}',t) + \right.$$
$$\left. + \int d^3\mathbf{r}' \int d^3\mathbf{p}' \sum_{n=0}^{\infty} \frac{1}{n!} \tilde{F}^{(2)}_{(\alpha_1\alpha_2...\alpha_n)}(\mathbf{r}-\mathbf{r}') \frac{\mathbf{p}'}{m} \frac{\partial}{\partial \mathbf{r}'} S_2^{(\alpha_1\alpha_2...\alpha_n)}(\mathbf{r},\mathbf{r}',\mathbf{p}',t) \right). \qquad (29)$$

This is the basic equation of the present paper. In a certain sense, in the expressions (25) and (29) the moments $S_l^{(\alpha_1\alpha_2...\alpha_n)}$ are the analogs of the distribution functions typical for the kinetic equations. Moreover, unlike the equations of motion of a system of material points [6], here it is important how the interaction potential and its derivatives behave in a physically infinitesimal volume. This determines the physical reason for the termination of the series on one or another term of the expansion. Moreover, in contrast to the case of kinetic equations, in the expressions containing $S_l^{(\alpha_1\alpha_2...\alpha_n)}$, the double summation is divided into two independent sums. This is due to the inclusion of Newton's third law for particles in the same volume $\Delta$.

4. **Particular cases of the general equation of motion**

If the internal fields vary slightly within physically infinitesimal volumes, then in the expansion (23), (24) we can leave the first few terms.

**A. First approximation**

We leave at first the moments of zero order:

$$S_1^{(0)}(\mathbf{r},\mathbf{r}',t) = \frac{1}{\Delta^2} \sum_{i=1}^{N} \sum_{k \neq i}^{N} \int_{\Delta(\mathbf{r})} d^3\xi \int_{\Delta(\mathbf{r}')} d^3\xi' \delta(\mathbf{r}+\boldsymbol{\xi}-\mathbf{r}_i(t)) \delta(\mathbf{r}'+\boldsymbol{\xi}'-\mathbf{r}_k(t)),$$

$$S_2^{(0)}(\mathbf{r},\mathbf{r}',\mathbf{p}',t) = \frac{1}{\Delta^2} \sum_{i=1}^{N} \sum_{k \neq i}^{N} \int_{\Delta(\mathbf{r})} d^3\xi \int_{\Delta(\mathbf{r}')} d^3\xi' \delta(\mathbf{r}'+\boldsymbol{\xi}'-\mathbf{r}_k(t)) \delta(\mathbf{r}+\boldsymbol{\xi}-\mathbf{r}_i(t)) \delta(\mathbf{p}'-\mathbf{p}_k(t)).$$

In this case, the expression for the volume density of the force takes the form:

$$\Phi(\mathbf{r},t) = -\int d^3\mathbf{r}' \frac{\partial U(\mathbf{r}-\mathbf{r}')}{\partial \mathbf{r}} S_1^{(0)}(\mathbf{r},\mathbf{r}',t) - \\ -\int d^3\mathbf{r}' \int d^3\mathbf{p}' \tilde{F}_{(0)}^{(2)}(\mathbf{r}-\mathbf{r}') \frac{\mathbf{p}'}{m} \frac{\partial}{\partial \mathbf{r}'} S_2^{(0)}(\mathbf{r},\mathbf{r}',\mathbf{p}',t). \tag{30}$$

Here

$$\tilde{F}_{(0)}^{(2)}(\mathbf{r}-\mathbf{r}') = \frac{\partial}{\partial \mathbf{r}}\left(U(\mathbf{r}-\mathbf{r}')\frac{|\mathbf{r}-\mathbf{r}'|}{c}\right) \equiv \frac{\partial}{\partial \mathbf{r}} U_{eff}(\mathbf{r}-\mathbf{r}'), \tag{31}$$

and $U_{eff}(\mathbf{r}-\mathbf{r}')$ is the effective renormalized interparticle potential. Taking into account the Newton's third law for the particles, we represent (30) in another form:

$$\Phi(\mathbf{r},t) = -n(\mathbf{r},t)\int d^3\mathbf{r}' \frac{\partial U(\mathbf{r}-\mathbf{r}')}{\partial \mathbf{r}} n(\mathbf{r}',t) - \\ -n(\mathbf{r},t)\int d^3\mathbf{r}' \int d^3\mathbf{p}' \frac{\partial U_{eff}(\mathbf{r}-\mathbf{r}')}{\partial \mathbf{r}} \frac{\mathbf{p}'}{m} \frac{\partial}{\partial \mathbf{r}'} f(\mathbf{r}',\mathbf{p}',t). \tag{32}$$

The introduced function

$$f(\mathbf{r}',\mathbf{p}',t) = \frac{1}{\Delta}\sum_k \int_{\Delta(\mathbf{r}')} d^3\xi' \delta(\mathbf{r}'+\xi'-\mathbf{r}_k(t))\delta(\mathbf{p}'-\mathbf{p}_k(t)) \tag{33}$$

is the result of smoothing by the microscopic distribution function (the so-called microscopic phase density) often used by Klimontovich [19,14]. We transform the integral in the second term on the right-hand side of (32):

$$\int d^3\mathbf{r}' \int d^3\mathbf{p}' \frac{\partial U_{eff}(\mathbf{r}-\mathbf{r}')}{\partial \mathbf{r}} \frac{\mathbf{p}'}{m} \frac{\partial}{\partial \mathbf{r}'} f(\mathbf{r}',\mathbf{p}',t) = \\ = \frac{1}{m}\int \frac{\partial U_{eff}(\mathbf{r}-\mathbf{r}')}{\partial \mathbf{r}} d^3\mathbf{r}' \frac{\partial}{\partial x'_\alpha} \frac{1}{\Delta}\int_{\Delta(\mathbf{r}')} \sum_k p'_{\alpha k}\delta(\mathbf{r}'+\xi'-\mathbf{r}_k(t))d^3\xi' = \\ = -\int \frac{\partial U_{eff}(\mathbf{r}-\mathbf{r}')}{\partial \mathbf{r}} \frac{\partial n(\mathbf{r}',t)}{\partial t} d^3\mathbf{r}'. \tag{34}$$

Then the equation for the local density takes the form:

$$\frac{\partial^2 n(\mathbf{r},t)}{\partial t^2} - \frac{1}{m}\nabla_\mathbf{r} \cdot \nabla_\mathbf{r} \mathbf{\Pi}(\mathbf{r},t) = \frac{1}{m}\nabla_\mathbf{r}\left(n(\mathbf{r},t)\nabla_\mathbf{r}\int d^3\mathbf{r}' U(\mathbf{r}-\mathbf{r}')n(\mathbf{r}',t)\right) - \\ -\frac{1}{m}\nabla_\mathbf{r}\left(n(\mathbf{r},t)\nabla_\mathbf{r}\int U_{eff}(\mathbf{r}-\mathbf{r}')\frac{\partial n(\mathbf{r}',t)}{\partial t}d^3\mathbf{r}'\right). \tag{35}$$

The resulting equation is noninvariant with respect to the time reversal operation. In addition to the local density $n(\mathbf{r},t)$ itself, the dynamics of the particle system is also determined by its changing rate $\partial n(\mathbf{r}',t)/\partial t$. If we neglect the retardation (i.e., at $c \to \infty$), the second term in the equation (35) becomes zero. The equation becomes invariant with respect to time reversal and, as a consequence, no longer describes the irreversible behavior of the particle system.

We also note that in the case of statistical averaging, the first term on the right-hand side of (35) can be expressed in terms of the functional derivative of the Helmholtz energy with respect to density. Then equation (35) becomes the basic equation of DDFT [16-18].

### B. Second approximation

If we take into account the first-order moments

$$S_1^{(\alpha)}(\mathbf{r},\mathbf{r}',t) = \frac{1}{\Delta^2}\sum_{i=1}^{N}\sum_{k\neq i}^{N} \int_{\Delta(\mathbf{r})} d^3\xi \int_{\Delta(\mathbf{r}')} d^3\xi' (\xi^\alpha - \xi'^\alpha)\delta(\mathbf{r}+\boldsymbol{\xi}-\mathbf{r}_i(t))\delta(\mathbf{r}'+\boldsymbol{\xi}'-\mathbf{r}_k(t)),$$

$$S_2^{(\alpha)}(\mathbf{r},\mathbf{r}',\mathbf{p}',t) = \frac{1}{\Delta^2}\sum_{i=1}^{N}\sum_{k\neq i}^{N} \int_{\Delta(\mathbf{r})} d^3\xi \int_{\Delta(\mathbf{r}')} d^3\xi' (\xi^\alpha - \xi'^\alpha)\delta(\mathbf{r}+\boldsymbol{\xi}-\mathbf{r}_i(t)) \times$$
$$\times \delta(\mathbf{r}'+\boldsymbol{\xi}'-\mathbf{r}_k(t))\delta(\mathbf{p}'-\mathbf{p}_k(t)),$$

then the expression for the volume density of force will take the form:

$$\begin{aligned}\boldsymbol{\Phi}(\mathbf{r},t) = &-n(\mathbf{r},t)\int d^3\mathbf{r}' \frac{\partial U(\mathbf{r}-\mathbf{r}')}{\partial \mathbf{r}} n(\mathbf{r}',t) + \\ &+n(\mathbf{r},t)\int \frac{\partial U_{eff}(\mathbf{r}-\mathbf{r}')}{\partial \mathbf{r}} \frac{\partial n(\mathbf{r}',t)}{\partial t} d^3\mathbf{r}' - \\ &-l^\alpha(\mathbf{r},t)\int d^3\mathbf{r}' n(\mathbf{r}',t)\frac{\partial}{\partial x^\alpha}\frac{\partial U(\mathbf{r}-\mathbf{r}')}{\partial \mathbf{r}} + n(\mathbf{r},t)\int d^3\mathbf{r}' l^\alpha(\mathbf{r}',t)\frac{\partial}{\partial x^\alpha}\frac{\partial U(\mathbf{r}-\mathbf{r}')}{\partial \mathbf{r}} + \\ &+l^\alpha(\mathbf{r},t)\int d^3\mathbf{r}' \tilde{F}_\alpha^{(2)}(\mathbf{r},\mathbf{r}')\frac{\partial n(\mathbf{r}',t)}{\partial t} - n(\mathbf{r},t)\int d^3\mathbf{r}' \tilde{F}_\alpha^{(2)}(\mathbf{r},\mathbf{r}')\frac{\partial l^\alpha(\mathbf{r}',t)}{\partial t}.\end{aligned} \tag{36}$$

Here

$$\tilde{F}_\alpha^{(2)}(\mathbf{r},\mathbf{r}') = \frac{|\mathbf{r}-\mathbf{r}'|}{c}\frac{\partial}{\partial x^\alpha}\left(\frac{\partial U(\mathbf{r}-\mathbf{r}')}{\partial \mathbf{r}}\right) + \frac{\partial}{\partial x^\alpha}\left(U(\mathbf{r}-\mathbf{r}')\frac{1}{c}\frac{\mathbf{r}-\mathbf{r}'}{|\mathbf{r}-\mathbf{r}'|}\right),$$

and $l^\alpha(\mathbf{r},t)$ is the vector field of the displacements of particles relative to the center of a physically infinitesimal volume:

$$l^\alpha(\mathbf{r},t) = \frac{1}{\Delta}\int_\Delta \sum_{i=1}^{N} \xi^\alpha \delta(\mathbf{r}+\boldsymbol{\xi}-\mathbf{r}_i(t)) d^3\xi. \tag{37}$$

Thus, in the second approximation, the contribution of the particle distribution in the volume $\Delta$, as well as the rate of change of this distribution $\partial l^\alpha(\mathbf{r}',t)/\partial t$, are taken into account for the volume density of the force. In the case of charged particles, the vector field (37) corresponds to an electric dipole moment.

### 5. Analysis of the equation of motion

We consider some consequences arising from the equation of motion (35), obtained in the first approximation with respect to the expansion of the potentials in the region of physically

infinitesimal volumes. To do this, we represent the tensor field of the momentum flux density in the form

$$\Pi_{\alpha\beta}(\mathbf{r},t) = mn(\mathbf{r},t)\mathrm{v}_\alpha(\mathbf{r},t)\mathrm{v}_\beta(\mathbf{r},t) + P_{\alpha\beta}(\mathbf{r},t), \quad (38)$$

where $P_{\alpha\beta}(\mathbf{r},t)$ is the tensor field of kinetic pressure:

$$P_{\alpha\beta}(\mathbf{r},t) = \frac{1}{\Delta}\int_\Delta \sum_{i=1}^{N} m u_{i\alpha}(t) u_{i\beta}(t) \delta(\mathbf{r}+\boldsymbol{\xi}-\mathbf{r}_i(t)) d^3\xi. \quad (39)$$

Here $\mathbf{u}_i(t) = \mathbf{v}_i(t) - \mathbf{v}(\mathbf{r},t)$ is thermal velocity of the $i$-th particle. We expand the tensor field $\mathbf{P}(\mathbf{r},t)$ onto the spherical and traceless parts:

$$\mathbf{P}(\mathbf{r},t) = \frac{1}{3}\mathbf{I}Sp\mathbf{P}(\mathbf{r},t) + dev\mathbf{P}(\mathbf{r},t), \quad (40)$$

where $\mathbf{I} = (\delta_{\alpha\beta})$ denotes the unit tensor, $Sp\mathbf{P}(\mathbf{r},t)$ is the trace of the tensor $\mathbf{P}(\mathbf{r},t)$, and $dev\mathbf{P}(\mathbf{r},t)$ denotes its deviator. With account of this, the equation of motion (35) can be represented in the form

$$\begin{aligned}\frac{\partial^2 n(\mathbf{r},t)}{\partial t^2} &= \nabla_\mathbf{r}\cdot\nabla_\mathbf{r}\left(n(\mathbf{r},t)\mathbf{v}(\mathbf{r},t)\otimes\mathbf{v}(\mathbf{r},t)\right) + \\ &+ \frac{1}{3m}\Delta\left(Sp\mathbf{P}(\mathbf{r},t)\right) + \frac{1}{m}\nabla_\mathbf{r}\cdot\nabla_\mathbf{r}dev\mathbf{P}(\mathbf{r},t) + \\ &+ \frac{1}{m}\nabla_\mathbf{r}\left(n(\mathbf{r},t)\nabla_\mathbf{r}\int d^3\mathbf{r}' U(\mathbf{r}-\mathbf{r}')n(\mathbf{r}',t)\right) - \\ &- \frac{1}{m}\nabla_\mathbf{r}\left(n(\mathbf{r},t)\nabla_\mathbf{r}\int U_{eff}(\mathbf{r}-\mathbf{r}')\frac{\partial n(\mathbf{r}',t)}{\partial t}d^3\mathbf{r}'\right).\end{aligned} \quad (41)$$

The trace of the kinetic pressure tensor is related to the energy density of the thermal motion of the particles $\rho(\mathbf{r},t)\varepsilon(\mathbf{r},t)$ and the scalar temperature field $T(\mathbf{r},t)$ by the following relation:

$$\begin{aligned}Sp\mathbf{P}(\mathbf{r},t) &= \frac{1}{\Delta}\int_\Delta \sum_{i=1}^{N} m u_i^2(t)\delta(\mathbf{r}+\boldsymbol{\xi}-\mathbf{r}_i(t))d^3\xi = \\ &= 2\rho(\mathbf{r},t)\varepsilon(\mathbf{r},t) = 3n(\mathbf{r},t)T(\mathbf{r},t).\end{aligned} \quad (42)$$

The concept of a local temperature as a smoothed energy of thermal motion of a physically infinitesimal volume arises naturally without the use of methods of statistical mechanics.

**A. The equilibrium state of a continuous medium**

Let us consider a continuous medium at a temperature $T$ placed in an external field $\varphi(\mathbf{r})$ in which there are no macroscopic flows of any type. The equation for the local density of the medium in this case takes the form:

$$T\Delta n(\mathbf{r}) = -\nabla_{\mathbf{r}} \cdot \nabla_{\mathbf{r}} dev\mathbf{P}(\mathbf{r}) - \\ -\nabla_{\mathbf{r}}\left(n(\mathbf{r})\left(\nabla_{\mathbf{r}}\int d^3\mathbf{r}' U(\mathbf{r}-\mathbf{r}')n(\mathbf{r}') + \nabla_{\mathbf{r}}\varphi(\mathbf{r})\right)\right). \tag{43}$$

In the case of local equilibrium $\nabla_{\mathbf{r}} \cdot \nabla_{\mathbf{r}} dev\mathbf{P}(\mathbf{r}) = 0$, and therefore equation (43) takes a simpler form:

$$T\Delta n(\mathbf{r}) = -\nabla_{\mathbf{r}}\left(n(\mathbf{r})\left(\nabla_{\mathbf{r}}\int d^3\mathbf{r}' U(\mathbf{r}-\mathbf{r}')n(\mathbf{r}') + \nabla_{\mathbf{r}}\varphi(\mathbf{r})\right)\right). \tag{44}$$

From the resulting equation, a solution of the Boltzmann type follows:

$$n(\mathbf{r}) = C\exp\left(-\frac{1}{T}\left(\int d^3\mathbf{r}' U(\mathbf{r}-\mathbf{r}')n(\mathbf{r}') + \varphi(\mathbf{r})\right)\right). \tag{45}$$

This equation was first obtained by A.A. Vlasov [20], starting from the concept of distribution functions for non-localized particles. Equation (45) can also be obtained in the framework of the density functional theory in the mean-field approximation [21]. Zakharov [6] recently obtained the same equation, but for a system of point particles with taking into account of its self-action.

**B. Weak nonequilibrium motion of a continuous medium**

Let us consider the isothermal dynamics of a continuous medium. We assume that the medium as a whole is at rest. We will also assume that there is no external field, and the local density of the medium undergoes small deviations $\tilde{n}(\mathbf{r},t)$ from some equilibrium value $n_0$:

$$n(\mathbf{r},t) = n_0 + \tilde{n}(\mathbf{r},t). \tag{46}$$

The velocities of physically infinitesimal volumes $\Delta$ are assumed to be small of the first order (like $\tilde{n}(\mathbf{r},t)$), which is a natural assumption for isothermal dynamics. Moreover, under this assumption, the condition of local equilibrium can be adopted. With these approximations, the linearized equation (41) takes the form:

$$\frac{\partial^2 \tilde{n}(\mathbf{r},t)}{\partial t^2} = \frac{T}{m}\Delta\tilde{n}(\mathbf{r},t) + \frac{n_0}{m}\Delta\int U(\mathbf{r}-\mathbf{r}')\tilde{n}(\mathbf{r},t)d^3\mathbf{r}' - \\ -\frac{n_0}{m}\Delta\int U_{eff}(\mathbf{r}-\mathbf{r}')\frac{\partial\tilde{n}(\mathbf{r},t)}{\partial t}d^3\mathbf{r}'. \tag{47}$$

It is noteworthy that an analogous equation can also be obtained for a system of point particles [6] without smoothing operation (2). However, in the case of the equation for the microscopic density obtained in [6], in general, the physical meaning of the local temperature is not clear. The temperature can be determined only for a certain set of point particles. In our case this is a physically infinitesimal volume. Moreover, in order to obtain an equation of the form (47) for microscopic density, it is necessary to include the self-interaction of the particles [6].

Representing $\tilde{n}(\mathbf{r},t)$ in the form of a Fourier integral

$$\tilde{n}(\mathbf{r},t) = \frac{1}{(2\pi)^3} \int \tilde{n}(\mathbf{k},t) \exp(i\mathbf{k}\mathbf{r}) d^3\mathbf{k},$$

it is not difficult to show that the characteristic equation for the Fourier transform $\tilde{n}(\mathbf{k},t)$ has the complex roots, which leads to damping or, conversely, to an amplification of the oscillations [6]. This in turn means that accounting the retarded interactions effects results in an irreversible evolution of the continuous medium.

## 6. Summary and conclusions

In this work, we demonstrate that it is possible to derive sequentially the equations of motion of a continuous medium, based on Newtonian mechanics, *without involving any probabilistic hypotheses*. A direct connection between the equations of motion at the microscopic and macroscopic levels makes it possible to describe the *irreversible* behavior of a continuous medium in the case of the *retardation of interactions* between the particles. The resulting equation of motion (29), which is irreversible in time for the local density of a continuous medium, indicates the feasibility of *physical foundation the thermodynamics* of macrosystems without involving the probability description inherent in statistical mechanics and kinetics.

In contrast to the kinetic equations, equation (29) does *not contain correlation functions*. It should be noted, however, that equation (29) is closed with respect to density *only* in the case of interparticle potentials (and their derivatives) that slightly changing over distances magnitude of order of $\sqrt[3]{\Delta}$. Otherwise, it is required to take into account the distribution of the particles inside the volumes $\Delta$ and the change rate of this distribution with respect to time. At that, the above mentioned distributions are described by the tensor quantities of the type (37) and their derivatives of the required order.

**Acknowledgments**

This work has been supported by the Ministry of Education and Science of Russian Federation within the framework of the project part of the state order (project No. 3.3572.2017/4.6).

The author grateful to Prof. A.Yu. Zakharov for useful discussions.